\documentclass[12pt,a4paper]{article}

\usepackage{epsfig}
\epsfxsize=\textwidth
\renewcommand{\includegraphics}[1]{\epsfbox{#1}}

\newcommand{\rat}[2]{{\textstyle\frac{#1}{#2}}}
\newcommand{\half}{{\rat12}}
\newcommand{\third}{{\rat13}}

\newcommand{\bK}{{\bf K}}
\newcommand{\bI}{{\bf I}}

\newcommand{\om}{h}

\newcommand{\bn}{{\bf n}}
\newcommand{\tn}{\tilde{n}}

\newcommand{\br}{{\bf r}}

\newcommand{\bQ}{{\bf Q}}

\newcommand{\bX}{{\bf X}}

\newcommand{\bu}{{\bf u}}

\newcommand{\te}{x}

\newcommand{\al}{\alpha}
\newcommand{\be}{\beta}
\newcommand{\de}{\delta}

\newcommand{\ep}{\epsilon}
\newcommand{\ka}{\kappa}
\newcommand{\tka}{\tilde{\kappa}}

\newcommand{\eq}[1]{\label{eq:#1}}
\newcommand{\re}[1]{(\ref{eq:#1})}

\newcommand{\bstress}{\mbox{\boldmath$\tau$}}

\newcommand{\bde}{{\bf e}}

\newcommand{\btau}{{\bf t}}
\newcommand{\nab}{\mbox{\boldmath$\nabla$}}

\newcommand{\cF}{{\cal F}}
\newcommand{\cS}{{\cal S}}

\newcommand{\cA}{{\cal A}}
\newcommand{\cR}{{\cal R}}
\newcommand{\D}[2]{\frac{\partial #1}{\partial #2}}
\newcommand{\DD}[2]{\frac{\partial^2 #1}{\partial #2^2}}
\newcommand{\Ord}[1]{{\cal O}\left(#1\right)}

\newcommand{\reduce}{{\sc reduce}}
\newcommand{\vel}{{\vec u}}

\newcommand{\grad}{\nab}
\newcommand{\divv}{\nab\cdot}

\newcommand{\grads}{\nab}
\newcommand{\divs}{\nab\cdot}

\newcommand{\vx}{\vec x}

\newcommand{\hsx}{{m_{1}}}
\newcommand{\hsz}{{m_{2}}}
\newcommand{\hsa}{{m_{\alpha}}}

\newcommand{\hy}{{h_{3}}}
\newcommand{\qa}{{u_{\alpha}}}
\newcommand{\qx}{{u_{1}}}
\newcommand{\qz}{{u_{2}}}
\newcommand{\qy}{{v}}
\newcommand{\ex}{{\vec e_{1}}}
\newcommand{\ez}{{\vec e_{2}}}
\newcommand{\ey}{{\vec e_{3}}}
\newcommand{\hhx}{{\tilde h_{1}}}
\newcommand{\hhz}{{\tilde h_{2}}}
\newcommand{\hha}{{\tilde h_{\alpha}}}
\newcommand{\kx}{{k_{1}}}
\newcommand{\kz}{{k_{2}}}
\newcommand{\kal}{{k_{\alpha}}}

\renewcommand{\vec}[1]{{\bf #1}}

\makeatletter
\def\@oddfoot{\hfil\footnotesize\sf Roy, Roberts \& Simpson, \today}
\makeatother

\begin{document}

\title{\sf A Lubrication Model of Coating Flows over a Curved
        Substrate in Space}
\author{\em
R.~Val\'{e}ry Roy\thanks{Dept of Mechanical Engineering, University of
Delaware, USA. E-mail: \texttt{roy@me.udel.edu}}
        \and\em
A.J.~Roberts\thanks{Dept of Mathematics \& Computing, University of
Southern Queensland, Toowoomba, Queensland 4352, Australia.  E-mail:
\texttt{aroberts@usq.edu.au}}
        \and\em
M.E.~Simpson\thanks{Dept of Mathematics \& Computing, University of
Southern Queensland, Toowoomba, Queensland 4352, Australia.  E-mail:
\texttt{simpsonm@usq.edu.au}}
          }
\date{April 23, 1997}
        \maketitle

\begin{abstract}
		Consider the three-dimensional flow of a viscous Newtonian 
		fluid upon an abitrarily curved substrate when the fluid film 
		is thin as occurs in many draining, coating and biological 
		flows.  We derive a model of the dynamics of the film, the 
		model being expressed in terms of the film thickness and the 
		curvature tensor of the substrate.  The model accurately 
		includes the effects of the curvature of the substrate, via a 
		physical multiple-scale approach, and gravity and inertia, via 
		more rigorous centre manifold techniques.  Numerical 
		simulations exhibit some generic features of the dynamics of 
		such thin fluid films on substrates with complex curvature.
\end{abstract}

\tableofcontents

\section{Introduction}

The importance of thin-film fluid flows in countless industrial and
natural processes has lead to the development of a variety of
mathematical models and numerical simulations.  This has increased the
understanding of the various complex processes at play, such as the
effects of generalized Newtonian and shear-thinning rheology of the
liquid \cite{Weidner94}, of geometric complexity of the substrate
\cite{Schwartz95a}, of multicomponent mixtures and drying processes
\cite{Cairn96}, of surface contamination \cite{Schwartz95b,Schwartz96}
and roughness \cite{Sweeney93}, of advecting and diffusing
contaminants, and of moving substrates.  See also for example
\cite{Chang94,Moriarty91,Moriarty92,Moriarty93,Ruschak85,Tuck90a,Weidner96}.
 Examples of industrial applications include the coating processes of
autobodies, beverage containers, sheet goods and films, decorative
coating, gravure roll coating, etc.  Physical applications can be
found in the biomedical field such as the liquid films covering the
cornea of the eye or protecting the linings of the lungs
\cite{Grotberg94}.

Herein we consider the slow motion of a thin liquid layer over an
arbitrarily curved solid substrate.  The fluid is assumed to be
incompressible and Newtonian, constituted of a single component and
uncontaminated by surface-active material.  The substrate is
stationary.  The effects of substrate curvature on the flow of thin
liquid layers driven by surface tension were first modeled by Schwartz
\& Weidner \cite{Schwartz95a} for two-dimensional geometries.
Short-wavelength irregularities of the fluid's free surface are
quickly leveled by surface tension forces, but the long-term evolution
of the flow is primarily determined by the substrate's curvature.
Their calculations confirm qualitative observations of the thinning of
coating layers at outside corners and of the thickening at inside
corners.  In a subsequent work \cite{Schwartz95b}, they studied the
joint effect of substrate curvature and the presence of surfactant and
showed that corner defects may be significantly altered by the
presence of Marangoni forces.

The model derived herein exploits the thinness of the fluid layer and
leads to a ``lubrication'' model of the dynamics whereby the unknown
fluid fields are parameterised by only the fluid layer's thickness.
Thus, a considerable reduction of the dimensionality of the problem is
achieved when compared to the full Navier-Stokes equations.  Such
thin-geometry or long-wave models have previously been derived in
many physical contexts, such as in the mechanics of beams, plates and
shells \cite{Timo59,Roberts93}, coating flows
\cite{Levich62,Roskes69,Benney66,Atherton76,Tuck90a}, shallow-water waves
\cite{Mei89}, viscous fluid sheets \cite{VandeF95}, etc.  On a flat
substrate the usual model for the evolution of a viscous film's
thickness $\eta$, driven only by surface tension, is given by the
following non-dimensional equation
\begin{displaymath}
        \D\eta{t}\approx -\third\divv\left[\eta^3\grad\tka\right]\,,
\end{displaymath}
where $\tka\approx\grad^2\eta$ is the curvature of the free surface of
the film.  Based upon the conservation of fluid and the Navier-Stokes
equations, outlined in \S\S\ref{sseqn}--\ref{ssbcs}, we derive in
\S\ref{sbase} the following model for the evolution of a film on a curved
substrate:
\begin{equation}
        \D\zeta{t}\approx -\third\divv\left[ \eta^{2}\zeta\grad\tka
        -\half\eta^{4}(\kappa \bI-\bK)\cdot\grad\kappa \right]\,,
        \label{initstate}
\end{equation}
where $\zeta= \eta-\half\kappa\eta^2 + \third k_{1} k_{2} \eta^{3}$
is proportional to the amount of fluid locally above the
substrate; $\bK$ is the curvature tensor of the substrate; $k_{1}$,
$k_{2}$, and $\kappa = k_{1}+k_{2}$ are the principal curvatures and
the mean curvature of the substrate, respectively; and the
$\nab$-operator is expressed in a coordinate system of the substrate,
as introduced in \S\ref{scoord}.  This model systematically accounts
for the curvature of the substrate and that of the surface of the
film.  It improves the model derived by Schwartz \& Weidner
\cite{Schwartz95a} and extends it to flows where the substrate
curvature has a larger effect on the fluid dynamics.

Modifications to this model are derived, in \S\ref{scent}, when 
gravity and/or inertia are significant influences.  The analysis is 
performed using computer algebra and is based on centre manifold 
theory \cite{Roberts88a} to assure us that all the dynamical effects 
are incorporated into the model.  In centre manifold theory competing 
small effects need not appear at leading order in the analysis; thus 
we obtain the flexibility to adapt the model to a variety of parameter 
regimes \emph{without} redoing the whole analysis.  For further 
discussion of this and other aspects of the application of centre 
manifold theory to low-dimensional modelling see the review in 
\cite{Roberts97a}.  Here in particular, the various regimes of 
gravitational forcing around a curved surface are all encompassed 
within the one model, namely equation~(\ref{geqn}).  The terms 
affected by fluid inertia appear very small in generic situations, 
see~(\ref{eqrey}), and perhaps may be best used to indicate the error 
in the lubrication models for moderate Reynolds numbers.

In the last section, \S\ref{snumic}, we conclude with some numerical
simulations of flows on curved surfaces, with attention given to the
quantitative differences found between our model and that obtained in
\cite{Schwartz95a}.

\subsection{Equations of motion}
\label{sseqn}

We solve the Navier-Stokes equations for an incompressible fluid of 
density $\rho$ and viscosity $\mu$ moving with velocity field $\vel$ 
and pressure field $p$.  The flow is primarily driven by pressure 
gradients along the substrate and caused by capillary forces 
characterized by surface tension $\sigma$ and varying due to 
variations of the curvature of the free surface of the fluid.  
However, there may also be a gravitational body force, $\vec g$, of 
magnitude $g$ in the direction of the unit vector $\hat{\vec g}$.  
Suppose the film has characteristic thickness $H$.  We 
non-dimensionalise the equations by scaling variables with respect to: 
the reference length $H$; the reference time $\mu H/\sigma$; the 
reference velocity $U=\sigma/\mu$, and the reference pressure 
$\sigma/H$.  Thus, in this non-dimensionalisation we take the view of 
a microscopic creature of a size comparable to the thickness of the 
fluid; later we require that both the substrate and the free surface 
curve only gently when viewed on this microscale.  The 
non-dimensional fluid equations are then
\begin{eqnarray}
        \divv\vel & = & 0\,,  \label{eqincomp}\\
        \cR\left[ \D\vel t+\vel\cdot\grad\vel\right] & = &
        -\grad p+\grad^2\vel +b\hat{\vec g}\,,\label{eqns}
\end{eqnarray}
where $\cR={\sigma\rho H}/{\mu^{2}}$ is effectively a Reynolds
number characterising the importance of the inertial terms---it may be
written as $UH/\nu$ for the above reference velocity---and $b=\rho g
H^2/\sigma$ is a Bond number characterising the importance of the
gravitational body force when compared with surface tension.

In \S\ref{sbase} we assume that the regime of the flow is
characterized by a very small value of the Reynolds number so that the
inertia term $\cR D \bu / Dt$ is neglected in comparison to the
viscous forces in the fluid:
\begin{displaymath}
        \cR=U H / \nu \ll 1 \,.
\end{displaymath}
This is the ``creeping flow'' assumption of lubrication.  Later, in
\S\ref{scent}, we reinstate inertia into the analysis and determine
its leading order effects on the dynamics.

\subsection{Boundary conditions}
\label{ssbcs}

Preliminary statements of the boundary conditions are as follows.
\begin{enumerate}
\item The fluid immediately in contact with the substrate does not
slip along the stationary substrate $\cS$, that is
\begin{equation}
        \bu=\vec 0\quad\mbox{on }\cS\,.
        \label{noslip}
\end{equation}

\item The kinematic boundary
condition at the free surface of the fluid states that fluid
particles must follow the free surface.

\item The free surface, denoted by $\cF$ and assumed free of
contamination, must have zero-shear (tangential stress), namely
\begin{equation}
        \tilde{\bstress}\cdot\tilde\btau_1=\tilde{\bstress}\cdot\tilde\btau_2=0
        \quad\mbox{on }\cF\,,
        \label{fsstress}
\end{equation}
where $\tilde{\bstress}$ is the deviatoric stress acting across $\cF$, and
$\tilde\btau_\alpha$ are independent tangent vectors to $\cF$.  This
condition assumes a light and inviscid medium (such as air) above the
fluid layer---a condition which guarantees the continuity of the
tangential stress as one passes across the interface $\cF$.

\item The normal surface stress at $\cF$ must account for the surface
tension whose effect is to create a discontinuity in the normal stress
proportional to the mean curvature of $\cF$: in non-dimensional form
\begin{equation}
        p=-\tka+\tilde{\bstress}\cdot\tilde\bn
        \quad\mbox{on }\cF\,,
        \label{fsnorm}
\end{equation}
where $p$ is the fluid pressure relative to the assumed zero pressure
of the medium above, and $\tilde\bn$ is the unit normal to $\cF$.
\end{enumerate}

We do not discuss boundary conditions at the lateral extremes of the 
substrate as the substrate is assumed to be so large in extent, when 
compared to the fluid thickness, that the dynamics of the fluid film 
are largely unaffected by the edge boundary conditions.  A rational 
method for deriving boundary conditions for nonlinear dynamical models 
such as~(\ref{initstate}) is explained in~\cite{Roberts92c}.

\section{The adopted coordinate system}
\label{scoord}

The flow is best analysed and solved in a coordinate system that
naturally fits the curving substrate.  Based upon the principal
directions of curvature of the substrate, we construct an orthogonal
curvilinear coordinate system in the neighbourhood of the substrate.
This natural coordinate system has some remarkable properties which
make a systematic analysis tractable.

\subsection{The curvilinear coordinate system}
\label{sssccs}

\begin{figure}[tbp]
        \centerline{ \epsfxsize=95mm\includegraphics{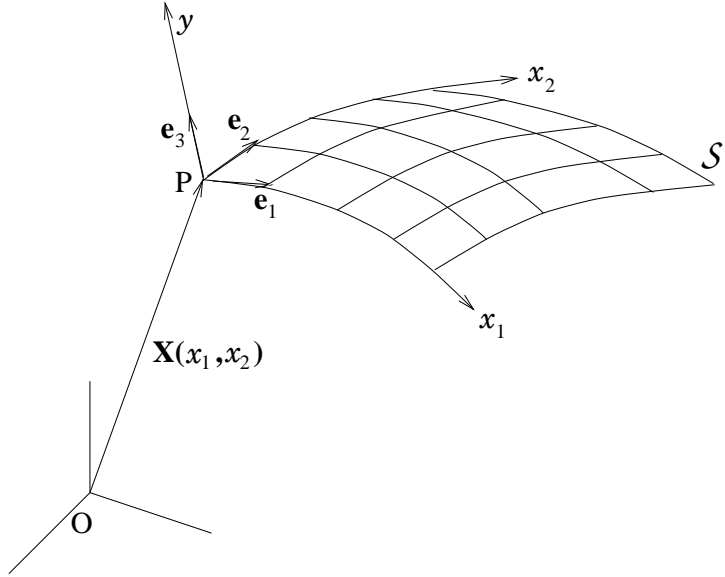}}
        \caption{The substrate $\cS$ is parameterised by variables $x_1$
        and $x_2$.  Together with the normal distance $y$, these form an
        orthogonal curvilinear coordinate system in space with unit
        vectors $\ex$, $\ez$ and $\ey$.  }
        \protect\label{fsurface}
\end{figure}

The curvilinear coordinate system is taken to be an extension into 
space of a natural coordinate system of the substrate.  We choose, 
without loss of generality, a curvilinear parameterisation $(x_{1}, 
x_{2})$ of $\cS$ for which the parameter curves $x_{1}= 
\mbox{Constant}$ and $x_{2}= \mbox{Constant}$ generate the lines of 
curvature of $\cS$ \cite{Stoker69}, shown schematically in 
Figure~\ref{fsurface}.  In this case, the curvature tensor becomes 
diagonal everywhere, with diagonal components $\kal$.\footnote{Latin 
indices (such as $i,j$) span the numbers $1,2,3$ and are attached to 
spatial quantities, while Greek indices (such as $\al$ or $\be$) span 
the numbers $1,2$ and are attached to substrate or surface quantities.  
A dash on Greek indices denotes the complementary value, that is, 
$\alpha'=3-\alpha$.  Subscripts following a comma indicate 
differentiation with respect to the corresponding coordinate.  Unless 
otherwise specified, we do \emph{not} use Einstein's summation 
convention for repeated indices.}

Conversely, we may view the surface $\cS$ as being entirely specified
by its metric coefficients $m_{\al} (x_{1},x_{2})$ and its
principal curvatures $k_{\al} (x_{1},x_{2})$ as functions of the
coordinate variables $(x_{1},x_{2})$ assumed to generate the lines
of curvature as the parameter curves of $\cS$.

We also define the triad of unit orthogonal vectors $(\bde_{1},
\bde_{2}, \bde_{3})$ at all points of $\cS$: $\bde_\alpha$ is tangent
to curves of constant $x_{\alpha'}$; and $\bde_3$ is normal to $\cS$.
The metric coefficients, $\hsa$, allow measurement of lengths on
$\cS$: the arclength of curves on the substrate are found from
\begin{equation}
d s ^{2}  = (m_{1} dx_{1})^{2} + (m_{2} dx_{2}) ^{2}\, .
\end{equation}
Note that the unit vectors vary along $\cS$ and that their derivatives
$\partial \bde_{i} / \partial \te_{\al}$ are needed to express the
equations of motion and boundary conditions in the curvilinear
coordinates.

As shown in Figure~\ref{fsurface}, the two-dimensional substrate $\cS$
is the locus of the endpoints of the position vector $ {\bf r}_{OP}=
\bX ( \te_{1} , \te_{2} )\,, $ for some domain ${\cal D}$ of
$(\te_1,\te_2)$.  We prescribe a third coordinate, denoted by either
$\te_{3}$ or $y$, as the distance measured along the normal $\bde_{3}$
from a given spatial point $P$ to the surface $\cS$.  Thus the
position vector of points $P$ of the fluid may be written as
\begin{equation}
\br = \bX (\te_{1} , \te_{2} ) + y \bde_{3} (\te_{1} , \te_{2} )\,,
\end{equation}
where the endpoint of vector $\bX$ belongs to the surface $\cS$.  A
large number of important simplifications occur in using this
particular orthogonal curvilinear coordinate system,
$(\te_1,\te_2,y)$, that naturally fits the substrate.  A definite
example for the flow on a torus is given in \S\S\ref{Storus}.

We denote by $\eta(t, \te_{1}, \te_{2})$ the fluid layer's thickness
at time $t$ and location $(\te_{1}, \te_{2})$.  The free surface $\cF$
of the fluid is thus represented by the equation $y = \eta(t, \te_{1},
\te_{2}) $, and the fluid fills the domain $ 0 \leq y \leq \eta(t,
\te_{1}, \te_{2})$.  Note that in general, sufficiently far from the
substrate for example, the chosen coordinate system does not lead to a
one-to-one mapping between coordinate space and physical space.  We
require that proper conditions between $\eta$, $k_{1}$ and $k_{2}$ are
satisfied so that intersections of normals of $\cS$ do not occur
within the body of the fluid.

The coordinate unit vectors are independent of $y$ and hence are the
same unit vectors, $(\bde_{1}, \bde_{2}, \bde_{3})$, as defined on the
substrate but now defined in space neighbouring $\cS$.  The
corresponding \emph{spatial} metric coefficients are
\begin{equation}
h_{\alpha}  = \hsa (1 - \kal y) \,, \quad
\hy = 1\,. \label{E:metric}
\end{equation}
By relating the base unit vectors of the rectangular coordinate system
to those of the curvilinear coordinate system, see
\cite[p598]{Batchelor67} or \cite{Morse53}, the spatial derivatives of
the curvilinear unit vectors are
\begin{equation}
\bde_{i,j} = { \bde_{i} \over \om_{i} }\om_{j,i} - \de_{ij}
\sum_{k=1}^{3} {\bde_{k} \over \om_{k}} \om_{i,k}\,.
\end{equation}
However, in the special coordinate system used here, all such
expressions simplify to the equivalent expressions on the substrate
and independent of $y$:
\begin{eqnarray}
&&
\bde_{\alpha,\alpha} =  -{m_{\alpha,\alpha'} \over m_{\alpha'} } \bde_{\alpha'}
+ k_{\alpha}m_{\alpha}\bde_{3}\,,\qquad
\bde_{\alpha,\alpha'} =  {m_{\alpha',\alpha} \over m_{\alpha} }
\bde_{\alpha'}\,,
\nonumber\\&&
\bde_{3,\alpha} =  -k_{\alpha} m_{\alpha} \bde_{\alpha}\,,\qquad
\bde_{i,3} = \vec 0\, .
\end{eqnarray}
This choice of curvilinear coordinates is needed only for the 
\emph{derivation} of a reduced-order approximation of the dynamics of 
coating flows on $\cS$.  As seen in (\ref{initstate}) for example, the 
model will be ultimately expressed in coordinate-free terms.

\subsection{Free-surface geometry}

The shape of the free surface is critical in thin film flows:
\emph{fluid surface} curvature variations create pressure gradients to
drive the fluid flow.  To denote a quantity evaluated on the fluid's
free surface we generally use an over-tilde (as already done in 
\S\S\ref{ssbcs}).

The position of points $P$ on the fluid's free surface $\cF$ is given
by
\begin{equation}
\br_{OP}= \widetilde{\bX}(\te_{1}, \te_{2}) = \bX (\te_{1}, \te_{2}) +
 \eta(t, \te_{1}, \te_{2}) \bde_{3}(\te_{1}, \te_{2})\,.
\end{equation}
Hence, the surface $\cF$ is naturally parameterised by
$({\te}_{1},{\te}_{2})$, and its tangential vectors, $\tilde{\btau}_{1}$
and $\tilde{\btau}_{2}$, and unit normal vector, $\tilde{\bn}$, are
given by
\begin{eqnarray}
\tilde{\btau}_{\alpha} &=& \D{\widetilde{\bX} }{\te_{\alpha} } =
\hha \bde_{\alpha} + \eta_{,\alpha} \bde_{3} \,,  \\
\tilde{\bn} &=& { \tilde{\btau}_{1} \times \tilde{\btau}_{2} \over |
\tilde{\btau}_{1} \times \tilde{\btau}_{2} | }
\propto -\hhz\eta_{,1}\ex-\hhx\eta_{,2}\ez+\hhx\hhz\ey
\,,\label{fnormal}
\end{eqnarray}
where $\hha=\hsa(1-\kal\eta)$ are the metric coefficients at the
free surface, and where $\eta_{,\al} = \partial \eta / \partial
\te_{\al}$.

At the free surface, $y = \eta$, the kinematic boundary condition must
be imposed, namely that fluid particles on the free surface remain on
it.  This leads to
\begin{equation}
\D{\eta}{t} = \qy
- {\qx \over \hhx} \D{\eta }{\te_{1}}
- {\qz \over \hhz} \D{\eta }{\te_{2}}
\quad\mbox{on }\cF\,.
\eq{kin_bc}
\end{equation}

To account for surface tension effects, we need to compute the free
surface mean curvature $\tilde{\ka}$ in terms of the substrate
principal curvatures and the film thickness $\eta$.  A tractable route
is to recognise that the effect of surface tension arises through the
energy stored in the free surface.  Thus its contribution to the
dynamical equations, through the curvature $\tka$, arises from the
variation of the surface area with respect to changes in the
free-surface shape $y=\eta(t,x_1,x_2)$.  The free-surface area is
$A=\int dA=\int \cA\, dx_1dx_2$ where
\begin{displaymath}
        \cA=\sqrt{\hhx^{2}\hhz^{2} +\hhz^{2}\eta_{,1}^{2}
        +\hhx^{2}\eta_{,2}^2}\,,
\end{displaymath}
is proportional to the free-surface area above a patch $m_1 dx_{1}
\times m_2 dx_{2}$ of the substrate.  The effect of curvature of the
free surface, $\tka$, is determined from the variation of $\cA$ with
respect to $\eta$:
\begin{displaymath}
        \hhx\hhz\tka = -\frac{\delta\cA}{\delta\eta}
        = \D{\ }{x_{1}}\left(\D\cA{\eta_{,1}}\right)
        +\D{\ }{x_{2}}\left(\D\cA{\eta_{,2}}\right)
        -\D\cA\eta \,.
\end{displaymath}
That is,
\begin{eqnarray}
        \tka&=&\frac{1}{\hhx\hhz}\left[
             \D{\ }{x_{1}}\left(\frac{\hhz^{2}\eta_{,1}}{\cA}\right)
           +\D{\ }{x_{2}}\left(\frac{\hhx^{2}\eta_{,2}}{\cA}\right)
           \right]
        \nonumber\\&&
        +\frac{1}{\cA}\left[
             \left(\hhx^{2}+\eta_{,1}^{2}\right)\frac{\hsz\kz}{\hhx}
          + \left(\hhz^{2}+\eta_{,2}^{2}\right)\frac{\hsx\kx}{\hhz}
          \right]\,.
\end{eqnarray}
An approximation is
        \begin{equation}
                \tka=\grads^2\eta +\frac{\kx}{1-\kx\eta}+\frac{\kz}{1-\kz\eta}
                +\Ord{\kappa^{3}+\nab^{3}\eta}\,,
                \label{fsc}
        \end{equation}
        where it is sufficiently accurate to use the standard form of the
        Laplacian in the curvilinear coordinates of the substrate,
        \begin{displaymath}
                \grads^2\eta=\frac{1}{\hsx\hsz}\left[ \D{\ }{x_{1}}
                \left(\frac{\hsz}{\hsx}\D\eta{x_{1}}\right) +\D{\ }{x_{2}}
                \left(\frac{\hsx}{\hsz}\D\eta{x_{2}}\right)\right]\,.
        \end{displaymath}
The approximation~(\ref{fsc}) arises directly from the form of the
variational expression and accounts for changes in the free-surface
curvature due to the finite depth of the film, and variations in the
film thickness.

Note that throughout this paper, $\grad$ has two different meanings 
depending upon the context of whether it is applied to 
three-dimensional spatial fields, such as $\vel$ and $p$, or to 
two-dimensional substrate fields such as $\eta$ and $\kappa$.

\section{Lubrication model driven by surface tension}
\label{sbase}

We now derive a model for the flow dynamics when the fluid film and 
the substrate vary on a large scale relative to the thickness of the 
film: because of the two vastly different space scales it may be 
viewed as a multiple-scale analysis.  In this case, viscous 
dissipation acts quickly across the fluid to damp out all except the 
slow dynamics associated with such large scale motion.  With the 
assumptions that inertia and body forces can be neglected, by 
considering conservation of mass we derive an equation for the {\em 
slow} evolution of the film thickness $\eta$.

\subsection{Re-scale the problem}
\label{ssscal}

We need to re-scale the non-dimensional governing equations.  There
are two characteristic lengths  of the problem: a
reference length $L$ measured along the substrate and a reference
thickness $H$ of the fluid layer covering $\cS$.  The length-scale $L$
is thought of as either the scale of the radius of curvature of the
substrate, or as the scale on which the film thickness varies.  We
generally expect both to be of the same order of magnitude as they
both are inversely proportional to substrate gradients, $\grads\bn$
and $\grads\eta$ respectively.  We denote the
ratio $H/L$ by $\ep$, and assume $\ep \ll 1$ to be consistent with the
thin-film/large-substrate assumption.  In particular, we consider the
flow to be non-dimensionally of thickness $1$, and so the substrate
scale is non-dimensionally of large size $1/\epsilon$.

With the above in mind, we introduce the scaled curvatures 
(temporarily indicated by a ``*'' superscript)
\begin{equation}
        k_{\al} = \epsilon k_{\al}^*,
        \quad \mbox{and }\tka=\epsilon\tka^*,
        \label{scalk}
\end{equation}
to express that the substrate and free-surface curvatures are
$\Ord{\epsilon}$.  Then we scale substrate coordinates and metric
coefficients according to
\begin{equation}
m_{\al}= \frac{1}{\epsilon}{m_{\al}^*}\,, \ \
\te_{\al}^{*} = \te_{\al}\,.
\label{scalm}
\end{equation}
This form is appropriate if the substrate coordinates are naturally
non-di\-men\-sion\-al, such as the angular latitude and longitude
coordinates on a sphere.  (If you consider the substrate coordinates
as naturally lengths, then the alternative scaling $m_{\al}=
{m_{\al}^*}$, $\te_{\al}^{*} =\epsilon \te_{\al}$ is appropriate; in
this case you would consider $\te_\al^*$ as a slow-space scale.)  The
re-scaled spatial metric coefficients then become
\begin{equation}
 \om_{\alpha}^{*} = m_{\alpha}^{*}(1- \ep k_{\alpha}^{*} y)\,.
\end{equation}
Seeking a slow flow leads to the following re-scaling of the
nondimensional variables for pressure, velocity, and
stress:
\begin{equation}
p= \epsilon p^{*}, \quad
u_\alpha  = \epsilon^2 u_\alpha^{*}, \quad
\qy = \ep^3 \qy^{*}, \quad
\bstress= \epsilon^2 \bstress^*\,.
\label{scalf}
\end{equation}

Now we write the equations and boundary condition in the scaled
variables.  In what follows, we drop the ``*'' superscript on all
re-scaled variables.  First, the continuity equation, $\divv\vel=\vec
0$,
\begin{equation}
\D{}{\te_{1}} ( \om_{2} \qx  ) +
\D{}{\te_{2}} ( \om_{1} \qz ) +
\D{}{y} ( \om_{1} \om_{2} \qy )
= 0\,,
\end{equation}
then, the Stokes momentum equation, $\grad p=\grad^2\vel$:
\begin{eqnarray}
&& {\bde_{1} \over \om_{1}}
\D{p }{\te_{1}}   +
 {\bde_{2} \over \om_{2} }
\D{p }{\te_{2}}
+ {1\over\ep} {\bde_{3}} \D{p }{y}
={1\over \om_{1}\om_{2}} \left\{
\epsilon^2\D{}{\te_{1}} \left( {\om_{2}\over \om_{1}}
\D{}{\te_{1}} \right)
+\right.\nonumber\\&&\quad\left.
+\epsilon^2 \D{}{\te_{2}} \left( {\om_{1}\over \om_{2}}
\D{}{\te_{2}} \right)
+ \D{}{y} \left( {\om_{1}\om_{2}}
{\partial\over \partial y} \right)\right\}
(\qx  \bde_{1}+ \qz \bde_{2} + \ep \qy \bde_{3} ) \,.
\end{eqnarray}
The boundary conditions now take the following forms.
\begin{itemize}
\item First, the no-slip boundary condition (\ref{noslip})  is
\begin{equation}
u_{i} = 0\,, \quad \mbox{on }y=0\,.
\end{equation}

\item Now consider the zero-shear-stress boundary condition at $y =
\eta$.  First, from~(\ref{fnormal}), express the components of the
normal unit vector $\tilde{\bn}$ in terms of the scaled variables:
\begin{equation}
c \tn_{\alpha} = - \ep (1 - \ep \eta k_{\alpha'} )
{\eta_{,\alpha} \over \hsa } \,,
\quad
c \tn_{3} = (1 - \ep \eta k_{1} ) (1 - \ep \eta k_{2} ) \,,
\end{equation}
where $c=|\tilde\btau_1 \times\tilde\btau_2|$ is the constant of
normalisation.  Furthermore, from~\cite[p599]{Batchelor67}, the
components of the non-dimensional (symmetric) deviatoric stress tensor,
$\bstress=(\nab\bu + \nab\bu^{T})$, become, upon scaling:
\begin{eqnarray}
\tau_{\alpha\alpha} & = &
2\epsilon\left[ {1 \over \hsa} \D\qa{x_{\alpha}} +
{m_{\alpha,\alpha'} \over m_{1}m_{2}} u_{\alpha'} \right]
+ \Ord{\ep^2}\,, \nonumber \\
\tau_{12} & = & \epsilon\left[ {1 \over m_{2}} \D\qx{x_{2}} + {1 \over
m_{1}} \D\qz{x_{1}} - {m_{1,2} \over m_{1} m_{2}} \qx - {m_{2,1} \over
m_{1}m_{2}} \qz\right] + \Ord{\ep^2}\,, \nonumber \\
                \tau_{\alpha 3} & = & \D\qa{y} + \epsilon \kal  \qa
                + \Ord{\ep^2}\,,        \label{scalstress} \\
                \tau_{33} & = &  2\epsilon\D\qy{y}\,.\nonumber
\end{eqnarray}
Thus to order $\ep$, the only contributing terms in the boundary
condition $\tilde{\bstress} \cdot \tilde{\btau}_{\alpha} =0$ is
$\tau_{\alpha 3} \tn_{3}$, namely
\begin{equation}
\D{\qa  }{y } + \ep\kal \qa  + \Ord{\ep^{2}} = 0\,.
\end{equation}

\item In order to write down the normal stress boundary condition on
the free surface, equation~(\ref{fsnorm}), we first need to express
the free-surface mean curvature $\tka$ to order $\ep$:
from~(\ref{fsc}) the scaled free-surface mean curvature is
\begin{displaymath}
\tka  =  \ka  + \ep \ka_{2} \eta + \ep  \grads^{2}\eta +
\Ord{\ep^{2}}\,,
\end{displaymath}
where $\ka=k_1+k_2$ and $\ka_{2} = k_{1}^{2} + k_{2}^{2}$.  Next, the
normal surface stress component to order $\ep^2$ in scaled form is
(using summation)
\begin{displaymath}
-p + \epsilon\tn_{i} \tn_{j} \tau_{ij} =
-p + 2 \epsilon^2\left(
\D{\qy }{y} - {\eta_{,1}\over m_{1}}
\D{\qx  }{y} - {\eta_{,2}\over m_{2}}
\D{\qz }{y} \right)
+ \Ord{\ep^{3}}\,.
\end{displaymath}
Hence we now write~(\ref{fsnorm}) as
\begin{equation}
p = - \ka - \ep \ka_{2} \eta - \ep \grads^{2}\eta  +
\Ord{\ep^{2}}
\,,\quad\mbox{on }y=\eta\,;
\end{equation}
viscous stresses have no influence on the normal stress to this
order in this scaling.
\end{itemize}

\subsection{Perturbation solution}

We now find a solution of these equations by assuming a perturbation
expansion of the unknown fields in terms of the small parameter $\ep$.
We write each component in the following expansion:
\begin{eqnarray}
u_\al &=&  u_\al^{(0)} + \ep u_\al^{(1)}  + \ep^{2} u_\al^{(2)} + \cdots
\,,  \nonumber \\
v&=&   v^{(0)} + \ep v^{(1)} + \ep^{2} v^{(2)} + \cdots \,,   \\
p&=&   p^{(0)} + \ep p^{(1)} + \ep^{2} p^{(2)} + \cdots   \,.
\nonumber
\end{eqnarray}
Then, at the leading order, we find  the following equations governing
$u_\al^{(0)}$, $v^{(0)}$ and $p^{(0)}$:
\begin{displaymath}
\D{p^{(0)} }{y } =  0\,,
\end{displaymath}
\begin{equation}
{\partial^{2} u_\al^{(0)} \over \partial y^{2} } =
{1\over m_\al} \D{p^{(0)} }{\te_\al } \,,
\end{equation}
\begin{displaymath}
\D{}{\te_{1}} ( m_{2} u_1^{(0)} ) +
\D{}{\te_{2}} ( m_{1} u_2^{(0)} ) +
m_{1}m_{2} \D{v^{(0)} }{y }
= 0\,,
\end{displaymath}
with the boundary conditions
\begin{displaymath}
u_\al^{(0)} = v^{(0)} = 0 \quad \mbox{at }  y = 0\,,
\end{displaymath}
\begin{equation}
\D{u_\al^{(0)} }{y }  = 0
\quad \mbox{at } y = \eta\,,
\end{equation}
\begin{displaymath}
p^{(0)} = -  \ka  \quad  \mbox{at } y = \eta\,.
\end{displaymath}
The solution of these equations is readily found to be the expected
locally parabolic flow driven by pressure gradients induced by
curvature variations:
\begin{displaymath}
p^{(0)} = - \ka\,,
\end{displaymath}
\begin{equation}
u_\al^{(0)}= - { 1 \over m_{\alpha}} \ka_{,\alpha}
\left(
\half y^{2} - \eta y
\right)\,,
\end{equation}
\begin{displaymath}
v^{(0)}= \grads^{2} \ka \left(
\rat{1}{6} y^{3} -  \half \eta y^{2}
\right) - \half \grads \ka \cdot \grads\eta\; y^{2}\,.
\end{displaymath}

At the next order of perturbation:
\begin{displaymath}
\D{p^{(1)} }{y } = 0\,,
\end{displaymath}
\begin{equation}
{\partial^{2} u_\al^{(1)}  \over \partial y^{2} }
- \ka \D{u_\al^{(0)} }{y }  = {1\over m_\al}
\left( \D{p^{(1)} }{\te_\al } + k_\al y
\D{p^{(0)} }{\te_\al }
\right)\,,
\end{equation}
\begin{displaymath}
\D{}{\te_{1}} ( m_{2} u_1^{(1)}  ) +
\D{}{\te_{2}} ( m_{1} u_2^{(1)} ) +
m_{1}m_{2} \D{v^{(1)} }{y }
= 0\,,
\end{displaymath}
with the boundary conditions
\begin{displaymath}
u_\al^{(1)} = v^{(1)} = 0 \quad \mbox{at } y = 0\,,
\end{displaymath}
\begin{equation}
\D{u_\al^{(1)}  }{y } = - k_\al u_\al^{(0)}
\quad \mbox{at } y=\eta\,,
\end{equation}
\begin{displaymath}
p^{(1)} = -  \left( \ka_{2} \eta +  \grads^{2}\eta \right)
\quad \mbox{at } y = \eta\,.
\end{displaymath}
We find
\begin{eqnarray}
p^{(1)} &=& -  \left( \ka_{2} \eta +  \grads^{2}\eta \right) \,,
\nonumber\\
u_\al^{(1)}  &=& -  {1\over m_{\alpha}} (\ka_{2} \eta + \grads^{2}\eta)_{,\alpha}
\left( \half y^{2} - \eta y \right)
- {k_\al\over 6m_\al} \ka_{,\alpha} y^{3}
\\&&\nonumber
-  {\ka \over m_\al} \ka_{,\alpha}
( \rat{1}{6} y^{3}
- \half \eta y^{2} + \half \eta^{2} y )\,,
\end{eqnarray}

\subsection{Conservation of mass}

The expressions in the previous subsection show how the velocity and
pressure fields respond to free-surface and substrate curvature.  Such
flow will thin the film in some places and thicken it in others.
Conservation of fluid then leads us to an expression for the evolution
of the film's thickness $\eta$ by the driven flow.  Initially we work
with non-dimensional but unscaled variables and coordinates before
returning to scaled quantities.
\begin{figure}[tbp]
       \centerline{\epsfxsize=95mm\includegraphics{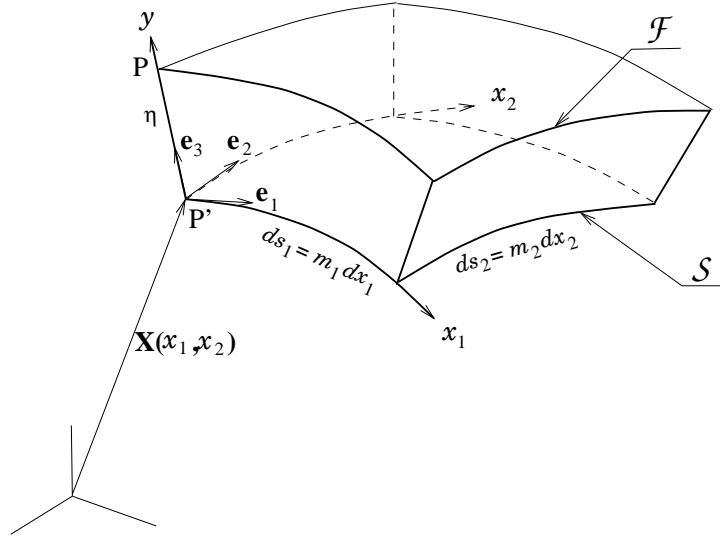}}
        \caption{A small control volume $V$ bounded by the substrate,
        $\cS$, the free surface, $\cF$, and four coordinate surfaces with
        separation $m_1dx_1$ and $m_2dx_2$.}
        \protect\label{fmasscons}
\end{figure}

Consider a small volume above a patch of the substrate extending
across the fluid layer from $y = 0$ to $y = \eta$, see
Figure~\ref{fmasscons}.  It is bounded by the substrate, the
instantaneous free surface and the coordinate surfaces $\te_{1}$, $\te_{1}
+ d\te_{1}$, $\te_{2}$, $\te_{2} + d\te_{2}$.  The rate at which fluid
leaves this volume is expressed to first order of infinitesimal
quantities $d\te_{1}$ and $d \te_{2}$ as the sum of the terms
\begin{equation}
\int_{0}^{\eta(\te_{1}+ d\te_{1}, \te_{2})}
\left[\qx  \om_{2} d\te_{2} \right]_{\te_{1}+ d\te_{1}} d y -
\int_{0}^{\eta(\te_{1}, \te_{2})} \left[ \qx  \om_{2} d\te_{2}
\right]_{\te_{1}} dy\,,
\end{equation}
for the surfaces $\te_{1}$ and $\te_{1} + d\te_{1}$, corresponding
terms for the surfaces $\te_{2}$ and $\te_{2} + d\te_{2}$, and the
term
\begin{equation}
\int_{\cF} \bu \cdot \widetilde{\bn} \,dS = \left\{
\hhx\hhz \qy
 - \eta _{,1} \hhz\qx  - \eta _{,2} \hhx \qz
\right\} d\te_{1}d\te_{2}
= \hhx\hhz \D{\eta  }{t} d\te_{1}d\te_{2}\,,
\end{equation}
for the bounding free surface $\cF$, where we have used the kinematic
boundary condition \re{kin_bc}.  Upon dividing by $d\te_{1} d\te_{2}$
and in the limit of $d\te_{1} d\te_{2} \to 0$, we deduce
\begin{equation}
m_{1}m_{2} (1- \eta  k_{1})(1-\eta  k_{2}) \D{\eta  }{t} =
- {\partial  \over \partial\te_{1} } (m_{2}  Q_{1} ) -
{\partial  \over \partial\te_{2} } (m_{1}  Q_{2} ) \,,
\label{Ealmost}
\end{equation}
where the $Q_{\al}$'s are the components of the total flux of fluid over
a position on the substrate, defined as
\begin{equation}
\bQ (t, \te_{1}, \te_{2}) = Q_{1} \bde_{1} + Q_{2} \bde_{2}
= \int_{0}^{\eta } ( (1 - k_{2}y) \qx  \bde_{1}
+ (1 - k_{1}y) \qz \bde_{2} ) dy\, .
\end{equation}
After division of~(\ref{Ealmost}) by $m_{1}m_{2}$, recognize on the
right-hand-side the surface divergence of the flux vector $\bQ$
expressed in the orthogonal curvilinear coordinate system $(\te_{1},
\te_{2})$ of $\cS$.  This yields the divergence form of the
conservation of mass equation:
\begin{equation}
(1- \eta  k_{1})(1-\eta  k_{2}) \D{\eta  }{t} =
-{1\over  m_{1} m_{2} } \left(
{\partial  \over \partial\te_{1} } (m_{2}  Q_{1} ) +
{\partial  \over \partial\te_{2} } (m_{1} Q_{2} )
\right)
= - \grads \cdot \bQ\,.
\end{equation}
We find an approximation of the flux vector $\bQ$ by taking into
account the thin layer characteristics of the flow over $\cS$ as
determined earlier.

Returning to the rescaled quantities introduced in~\S\S\ref{ssscal}, we now
determine the perturbation coefficients of the flux vector $\bQ =
\ep^{2}\bQ^{(0)} + \ep^3 \bQ^{(1)} + \cdots$ to be
\begin{equation}
\bQ^{(0)} = \int_{0}^{\eta } (u_1^{(0)} \bde_{1} + u_2^{(0)} \bde_{2} ) dy =
\rat{1}{3}  \eta ^{3} \left( {\bde_{1}\over m_{1}} \ka_{,1}
+ {\bde_{2}\over m_{2}} \ka_{,2} \right)
 =\rat{1}{3} \eta ^{3} \grads \ka\,,
\end{equation}
\begin{eqnarray}
\bQ^{(1)} &=& \int_{0}^{\eta } \left(
(u_1^{(1)}  - k_{2}y u_1^{(0)}) \bde_{1} + (u_2^{(1)}- k_{1}y u_2^{(0)})
\bde_{2}
\right) dy
\nonumber \\ &=&
 \rat{1}{3} \eta ^{3} \grads (\ka_{2} \eta  + \nab^{2}\eta )
+\rat{1}{6} \eta ^{4}
\bK \cdot \grads \ka - \rat{1}{3}  \eta ^{4} \ka \grads \ka\,,
\end{eqnarray}
where $\ka = k_{1}+k_{2}$, $\ka_{2}=k_{1}^{2}+k_{2}^{2}$, and
$\bK = k_{\al} \de_{\al \be} (\bde_{\al} : \bde_{\be})$ is the curvature
tensor (which would not be diagonal in a general orthogonal coordinate
system of the substrate $\cS$).

Finally, the corresponding evolution equation for $\eta $, the
so-called lubrication approximation, is given by
\begin{eqnarray}
(1- \ep \eta  k_{1})(1-\ep \eta  k_{2}) \D{\eta  }{t} &=&
-  \rat{1}{3} \ep^{3} \grads \cdot \left[
\eta ^{3} \grads\tka
-  \ep \eta ^{4} \ka \grads\ka
+\half \ep \eta ^{4} \bK \cdot \grads\ka
\right]
\nonumber\\&&\quad
+ \Ord{\ep^{5}}\,,
\end{eqnarray}
where $\tka = \ka + \ep \ka_{2} \eta  + \ep \grads^{2}\eta
+\Ord{\epsilon^2} $.

We express this equation in the more convenient form
\begin{equation}
\D{\zeta }{t} =
-  \rat{1}{3} \ep^{3} \grads \cdot \left[
\eta ^{2}\zeta \grads\tka
-\half \ep \eta ^{4}(\ka \bI - \bK) \cdot \grads\ka
\right]
+ \Ord{\ep^{5}}\,,  \label{lubeps}
\end{equation}
where $\zeta = \eta - \half \ep \ka \eta ^{2} + \rat{1}{3} \ep^{2}
k_{1}k_{2} \eta ^{3}$ is proportional to the amount of fluid in the
film lying ``above'' a patch of the substrate.  Rewriting this in
terms of unscaled, non-dimensional quantities gives~(\ref{initstate})
when the free-surface curvature is approximated by an expression such
as~(\ref{fsc}).

\section{Centre manifold analysis: Gravity and inertia}
\label{scent}

The previous analysis is adequate to give a basic model for the 
dynamics of the film thickness.  However, it is not clear how to 
reasonably modify such an analysis for generally curved substrates in 
the presence of more complicating physical effects such as gravity.  
In this section we turn to the more powerful centre manifold theory to 
extend the previous model in order to include a gravitational body 
force and to determine the leading effect of inertia.  The systematic 
nature of the centre manifold approach lends itself well towards 
computer algebra and hence its results provide a valuable check on the 
human algebra reported in the previous section.  In future work, this 
approach will lead to the correct modelling of initial conditions and 
forcing, as has been done for dispersion in pipes and channels 
\cite[e.g.]{Mercer90}, and perhaps small substrate roughness---see the 
review in \cite{Roberts97a} for an introduction.  This section also 
serves as an example of the analysis that may be undertaken if other 
physical effects are needed in the lubrication model of the dynamics.

\subsection{Basis of the centre manifold}

Centre manifold theory \cite{Carr81} assures us that the long-term
dynamics near a fixed point of a dynamical system may be accurately
modelled by a ``low-dimensional'' system which is based upon the
linear picture of the dynamics, see the review \cite{Roberts97a}.
Here the lubrication model~(\ref{initstate}) is of low-dimension in
comparison to that of the incompressible Navier-Stokes
equations~(\ref{eqincomp}--\ref{eqns}).  The first task is to
establish the linear basis of the centre manifold.

Analogous to the approach developed by Roberts
in~\cite{Roberts94c,Roberts96c}, we retain the small parameter
$\epsilon$ introduced in \S\ref{ssscal} and scale the curvatures and
metric coefficients according to~(\ref{scalk}--\ref{scalm}).  By
considering $\epsilon$ negligible we, in the preliminary \emph{linear
analysis}, restrict attention to large-scale flows on a flat
substrate.  However, we do not explicitly scale the unknown fields,
$p$ and $\vel$ as in (\ref{scalf}), because in this approach their
scaling naturally arises from the dynamical equations during the
course of the analysis rather than being imposed at the
outset---another virtue of this approach.  Further, we treat the
gravitational forcing as a small effect.  By substituting, as in
\cite{Roberts96c}, the Bond number $b=\beta^2$, then adjoining the
trivial equations
\begin{equation}
        \D\epsilon t=0\,,
        \quad\mbox{and}\quad
        \D\beta t=0\,,
        \label{triveb}
\end{equation}
to the scaled versions of the fluid
equations~(\ref{eqincomp}--\ref{eqns}) and their boundary conditions,
we treat all terms involving products of $\vel$, $\epsilon$ and
$\beta$ as perturbing nonlinear terms (as in the centre manifold
analysis of the unfolding of bifurcations \cite{Carr81}).

Such nonlinear terms are discarded in the preliminary linear analysis 
to leave linear equations
\begin{equation}
        \divv\vel=0\,,
        \quad\mbox{and}\quad
        \cR\D\vel t+\grad p-\grad^2\vel=\vec 0\,,
        \label{linns}
\end{equation}
where, since $\epsilon=0$, in effect these differential operators are
the Cartesian operators appropriate to a flat substrate.  These
equations are to be solved with boundary conditions of $\vel=\vec 0$
on $\cS$ ($y=0$) and
\begin{equation}
        \D\qa y=p-2\D\qy y=0
        \quad\mbox{on $y=\eta$}\,,
        \label{linbc}
\end{equation}
and the linearized kinematic condition
\begin{displaymath}
        \D\eta t-v=0\,,\quad\mbox{on $y=\eta$}\,.
\end{displaymath}
All solutions of these linear equations are composed of the decaying 
lateral shear modes, $\qy=p=0$, $\qa=c_\alpha\sin(\ell\pi 
y/(2\eta))\exp(\lambda t)$ for odd integers $\ell$, together with 
$\eta=\mbox{constant}$, $\qa=\qy=p=0$.  In these modes the decay-rate 
in time is $\lambda=-\ell^2\pi^2/(4\eta^2\cR)$, except for the last 
mentioned mode which, as a consequence of fluid conservation, has a 
decay-rate $\lambda=0$.  So linearly, and in the absence of any 
lateral variations on a flat substrate, all the lateral shear modes 
decay exponentially quickly, on a time-scale of $\cR\eta^2$, just 
leaving a film of constant thickness as the permanent mode.  This 
spectrum, of all eigenvalues being strictly negative except for a few 
that are zero, is the classic spectrum for the application of centre 
manifold theory; the Existence Theorem~1 in \cite{Carr81} assures us 
that the nonlinear effects just perturb the linear picture of the 
dynamics so that in the long-term all solutions of the full nonlinear 
system are dominated by the slow dynamics induced by gravity and 
large-scale lateral variations in the film thickness and curvature.  
The Relevance Theorem~2 in \cite{Carr81} assures us that these 
dynamics are exponentially attractive and so form a generic model of 
the long-term fluid dynamics of the film.  With the caveat that strict 
theory has not yet been extended sufficiently to cover this particular 
application, the closest being that of Gallay \cite{Gallay93} and 
Haragus \cite{Haragus95} (but also see \cite{Roberts88a}), we apply 
the centre manifold concepts and techniques to systematically develop 
a low-dimensional lubrication model of the dynamics of the film.

Having identified the critical mode associated with the zero
decay-rate, the subsequent analysis is straightforward.  The usual
approach is to write the fluid fields $\vec v(t)=(\qx,\qz,\qy,p)$, as
a function of the critical mode $\eta$ (effectively equivalent to the
``slaving'' principle of synergetics \cite{Haken83}).  Instead of
seeking asymptotic expansions in the ``amplitude'' of the critical
mode \cite[e.g.]{Roberts88a,Mercer90,Roberts94c}, we apply an
algorithm to find the centre manifold and the evolution thereon which
is based directly upon the Approximation Theorem~3 in
\cite{Carr81,Roberts96a} and its variants, as explained in detail by
Roberts in \cite{Roberts96a}.  An outline of the procedure follows.

We seek solutions for the fluid fields as
\begin{equation}
        \vec v(t)=\vec V(\eta)\,,
        \quad\mbox{such that}\quad
        \D{\eta}t=G\left(\eta,\vec V(\eta)\right)\,,
        \label{cmsub}
\end{equation}
where dependence upon the constant parameters $(\epsilon,\beta)$ is
implicit, and where $G$ is the right-hand side of the re-scaled
version of the kinematic condition~(\ref{eq:kin_bc}).  The aim is to
find the functional $\vec V$ such that $\vec v(t)$ as described
by~(\ref{cmsub}) forms actual solutions of the scaled Navier-Stokes
equations; this ensures fidelity between the model and the fluid
dynamics.  Suppose that at some stage in an iteration scheme we have
an approximation, $\tilde{\vec V}$ and $\tilde{G}=G(\eta,\tilde{\vec
V})$, and then seek a correction, $\vec V'$, so that we obtain a more
accurate solution to the governing equations.  Substituting
\begin{displaymath}
        \vec v(t)=\tilde{\vec V}+\vec V'\,,
        \quad\mbox{such that}\quad
        \D\eta t=\tilde{G}
\end{displaymath}
into the scaled Navier-Stokes equations then rearranging, dropping
products of corrections, and using a zeroth-order approximation
wherever factors multiply corrections, we obtain a system of equations
for the corrections which is of the form
\begin{equation}
        B\vec V'=\tilde{\vec R}\,,
        \label{itform}
\end{equation}
where $B$ is the linear operator seen on the left-hand side
of~(\ref{linns}--\ref{linbc}), and, most importantly, $\tilde{\vec R}$
is the residual of the scaled Navier-Stokes equations using the
current approximation.  After solving this equation to find the
correction $\vec V'$, the current approximation $\tilde{\vec V}$ and
$\tilde G$ is updated.  The iteration is repeated until the residual
of the governing equations, $\tilde{\vec R}$, becomes zero to some
order of error, whence the centre manifold model will be accurate to
the same order of error (by the Approximation Theorem~3 in \cite{Carr81}).

A computer algebra program\footnote{The computer algebra package
{\reduce} was used because of its flexible ``operator'' facility.  The
source code is publicly available via the URL
\texttt{http://\-www.\-sci.\-usq.\-edu.\-au/\-\~{}robertsa} or by
contacting the second listed author.  At the time of writing,
information about {\reduce} was available from Anthony C.\ Hearn,
RAND, Santa Monica, CA~90407-2138, USA. E-mail: \tt reduce@rand.org}
was written to perform all the necessary detailed algebra for this
physical problem.  A very important feature of this iteration scheme
is that it is performed until the residuals of the actual governing
equations are zero, to some order of error.  Thus the correctness of
the results that we present here is based only upon the correct
evaluation of the residuals and upon sufficient iterations to drive
these to zero.  Thus the key to the correctness of the results
produced by the computer program is the proper coding of the fluid
dynamical equations in the chosen curvilinear coordinate system.  Upon
obtaining the code, these can be seen in the computed residuals within
the iterative loop.  Also note that because the thickness of the film
is continuously varying in space and time, it is convenient to work
with equations in terms of a scaled vertical coordinate
$Y=y/\eta(\vx,t)$ so that the free surface of the film is always
$Y=1$.  We also let $X_{\alpha}=x_{\alpha}$, $T=t$ complete the new
coordinate system.  Then space and time derivatives transform
according to
\begin{displaymath}
        \D\ {y} = \frac{1}{\eta}\D\ Y\,, \quad
        \D\ {x_{\alpha}} = \D\
{X_{\alpha}}-\frac{Y}{\eta}\D\eta{X_{\alpha}}\D\ Y\,,  \quad
        \D\ t = \D\ {T}-\frac{Y}{\eta}\D\eta{T}\D\ Y\,.
\end{displaymath}
However, the fluid equations are not rewritten in this new coordinate
system because the computer handles all the necessary details of the
transformation.

\subsection{The general lubrication model}

Based upon scalings introduced in \S 3.1, we run the {\reduce} computer algebra
program that computes velocity and pressure fields for the flow and the
evolution equation for the film's thickness.
We find that the evolution
equation may be written in the coordinate-free
form\footnote{$\Ord{\epsilon^p+b^q}$ is used to denote terms, $z$, for
which $z/(\epsilon^p+b^q)$ is bounded as $(\epsilon,b)\to\vec 0$.  The
upshot is that $z=\epsilon^mb^n$ is $\Ord{\epsilon^p+b^q}$ iff
$m/p+n/q\geq 1$.  For example, an expression accurate to
$\Ord{\epsilon^{5}+b^{2}}$ retains all terms of the form
$\epsilon^mb^n$ for $2m+5n<10$.}
\begin{eqnarray}
        \D\zeta t&=&-\rat{1}{3}\epsilon^3\divs\left[ \eta^{2}\zeta\grads\tka
        -\rat{1}{2}\epsilon\eta^{4}(\kappa \bI-\bK)\cdot\grads\kappa \right]
        \nonumber \\&&
        -\rat{1}{3}\epsilon b\divs\left[\eta^{3}\hat{\vec g}_{s}
        -\epsilon\eta^{4}(\kappa \bI+\half \bK)\cdot\hat{\vec g}_{s}
        +\epsilon\hat g_{y}\eta^{3}\grads \eta \right]
        \nonumber \\&&
        +\Ord{\epsilon^{5}+b^{2}}\,,
        \label{geqn}
\end{eqnarray}
where $\hat{\vec g}_s$ and $\hat g_y$ are respectively the components
of the gravitational unit vector tangent and normal to the substrate.
Note that in the special orthogonal curvilinear system used in the
derivation of the model,
\begin{displaymath}
        \kappa \bI-\bK=\left[
        \begin{array}{cc}
                k_{2} & 0  \\
                0 & k_{1}
        \end{array}
        \right]\,,
        \quad\mbox{and}\quad
        \kappa \bI+\half \bK=\left[
        \begin{array}{cc}
                \rat32 k_1+k_{2} & 0  \\
                0 & k_1+\rat32 k_2
        \end{array}
        \right]\,.
\end{displaymath}

One example of the application of the model~(\ref{geqn}) is to the 
dynamics of thin films on vertical cylinders or fibres as explored by 
Kalliadasis \& Chang \cite{Kalliadasis94}.  They find that a thin film 
may form into drops on the fibre, or may saturate into solitary waves 
depending upon the film thickness.  The model they use, derived by 
Frenkel \cite{Frenkel92}, is approximately the version of~(\ref{geqn}) 
specific to axisymmetric flow around a cylinder of radius $a$, namely 
\begin{equation}
	\D\zeta t +\frac{1}{3}\D{\ }z\left[g\eta^3\left(1+\frac{\eta}{a}\right)
	+\eta^2\zeta \left(\frac{1}{(a+\eta)^2}\D\eta 
	z+\frac{\partial^3\eta}{\partial z^3}\right)\right]=0\,,
	\label{Ecyl}
\end{equation}
in our nondimensionalisation, where $z$ measures axial distance and 
here $\zeta=\eta+\eta^2/2a$.  The differences between our systematic 
approach and theirs is that their model, (2) in \cite{Kalliadasis94}, 
does not conserve fluid, while our model~(\ref{Ecyl}) does and also 
has some higher order corrections to account better for the curvature 
of the substrate.  Further, (\ref{geqn}) specialised to a cylindrical 
substrate will also descibe any non-axisymmetric dynamics of interest.

Returning to the general model~(\ref{geqn}), it is derived solely 
under the assumptions that curvature and free-surface slopes, as 
measured by $\epsilon$, and the gravitational forcing, measured by the 
Bond number $b$, are perturbing influences.  The application of centre 
manifold theory places no restrictions upon their relative 
magnitudes---we do \emph{not} have to insist on $b\sim\epsilon$ or any 
other such relation between these two independent parameters.  
Provided there are no ``run away'' instabilities, the model is valid 
over any scaling regime where both parameters are small.  In 
particular, the model is valid as the tangential and normal 
gravitational forcing vary widely around a curving substrate.

Also note that the model~(\ref{geqn}) was derived without placing any 
overt restriction upon the Reynolds number, it was treated as an 
$\Ord{1}$ constant.  That the model actually turns out to have no 
Reynolds number dependence just confirms the Stokes flow nature of 
these lubrication dynamics.  Higher-order analysis, as also seen in 
\cite[\S4]{Roberts96c}, shows that inertia first appears at 
$\Ord{\epsilon^6+b^3}$ for fluid films.  Thus inertia is formally 
negligible.  However, if the Reynolds number is large enough to be 
significant, then we have determined that the following 
$\Ord{\cR\epsilon^6+\cR b^3}$ terms should be included in the 
evolution equation~(\ref{geqn}):
\begin{eqnarray}
        \D\zeta t&=&\cdots\nonumber \\&&
        -\frac{\cR}{5}\divs\left[
        \frac{2\eta^6}{3}\left(\grads\eta\cdot\grads\kappa\right) \grads\kappa
        -\frac{\eta^7}{7}\grads \left(\grads\kappa\cdot \grads\kappa \right)
        +\frac{26\eta^7}{63}(\grads^2\kappa)\grads\kappa
        \right]
        \nonumber \\&&
        -\frac{2\cR b}{15}\divs\left[
        {\eta^6}\left( \hat{\vec g}_s\cdot\grads\eta\, \grads\kappa +
        \hat{\vec g}_s\,\grads\eta\cdot\grads\kappa \right)
        -\frac{3\eta^7}{7}\grads\left( \hat{\vec g}_s\cdot\grads\kappa \right)
        \nonumber\right.\\&&\qquad\left.
        +\frac{13\eta^7}{21}\hat{\vec g}_s\grads^2\kappa
        +\frac{13\eta^7}{21}\hat{g}_y\kappa\grads\kappa
        \right]
         \nonumber\\&&
        -\frac{2\cR b^2}{15}\divs\left[
        {\eta^6}\hat{\vec g}_s\left( \hat{\vec g}_s\cdot\grads\eta \right)
        +\frac{\eta^7}{21}(13\kappa \bI-9\bK)\cdot\hat{\vec g}_s \hat{g}_y
        \right]\,. \label{eqrey}
\end{eqnarray}
Note that for a flat substrate $(k_{1}=k_{2}=0)$, these Reynolds
number correction terms reduce to those derived by Benney~\cite{Benney66} and
Atherton \& Homsy~\cite{Atherton76}.
We suggest that these terms are more likely to be used to
estimate the error in the lubrication model~(\ref{geqn}) when applied
to moderate Reynolds number flows.  Thus these expressions may be used
to indicate when a more sophisticated dynamical model, such as a
two-mode model~\cite{Chang94,Roberts94c}, is required in order to
resolve the inertial instabilities of fluid films at higher Reynolds
numbers.  Also recall from the previous subsection that the time-scale
of decay of the lateral shear modes is $\cR \eta^2$ and so the other
criterion for validity of the model is that the time-scales exhibited
in a simulation are significantly longer than this.

Of course, an order of magnitude argument will give a global estimate,
over space and time, of the influence of inertia.  However, one may
make a perfectly satisfactory a priori order of magnitude assessment,
but if in a simulation an instability grows, then (\ref{eqrey}) will
detect if it grows too large.  Thus (\ref{eqrey}) supplements order of
magnitude estimates by giving focussed \emph{local} estimates of the
errors in an actual simulation.

\section{Example film flows}
\label{snumic}

In this section we use the preceding lubrication models to demonstrate
some of the flow effects caused by substrate curvature.

\subsection{Qualitative effect of substrate curvature}

\begin{figure}[tbp]
        \centerline{\includegraphics{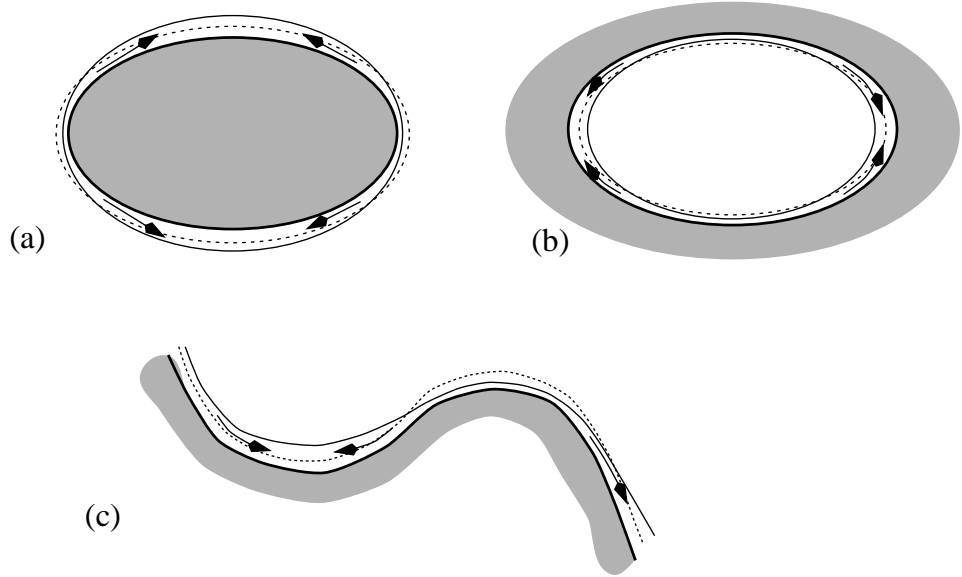}}
        \caption{Leading effects of substrate curvature on the evolution of
a thin
flow}
        \protect\label{fcorner}
\end{figure}

Equation~(\ref{lubeps}) shows that, to leading order of perturbation,
the flow is driven by substrate curvature gradients, curvature caused
by film thickness variations is generally smaller unless the substrate
is comparatively gently curved.  To understand the effect of curvature
qualitatively, consider a two-dimensional flow on a one-dimensional
substrate with given curvature $\kappa$.  Denote $x= x_{1}$ as substrate
arclength so that the scale factor $m_1=1$ and there are no variations
in $x_2$.  Then to leading order and in the absence of body forces,
the flow is governed by the first-order partial differential equation
for $\eta$:
\begin{equation}
(1- \ka \eta) \D{\eta}{t} = - \third \D\ x(\eta^{3}\ka_{x})\,.
\end{equation}
This has a characteristic solution
\begin{equation}
\dot{x} = \ka_{x} {\eta^{2} \over 1- \ka \eta}\,,\qquad
\dot{\eta} = - \third \ka_{xx} {\eta^{3} \over 1- \ka \eta}\,.
\label{charsol}
\end{equation}
\begin{itemize}
        \item Wherever the substrate curvature $\ka$ has a local minimum,
        say at $x=x_{0}$, then
\begin{displaymath}
\ka_{x} < 0  \  (x<x_{0})\,, \quad \ka_{x} >0 \  (x>x_{0})\,, \quad
\mbox{and}\quad \ka_{xx} > 0\,.
\end{displaymath}
Thus according to the characteristic equations the flow is thinned in
the neighborhood of $x=x_{0}$.  Indeed, since $\dot\eta\propto
-\kappa_{xx}\eta^3$, the film thickness $\eta$ typically decreases as
$t^{-1/2}$ at a point of minimum curvature $(\ka_{xx} >0)$.  As shown
schematically in Figure~\ref{fcorner}, this thinning applies in the
neighbourhood of minimum absolute curvature for interior coating
flows, and of maximum absolute curvature for exterior coating flows.

\item Conversely, the film thickens in the neighbourhood of a point of
maximum curvature as also shown in Figure~\ref{fcorner}.  The
characteristic solution~(\ref{charsol}) predicts a film thickness that
blows up in finite time.  The role of higher-order terms in the model
equations is to smooth out such singularity.

\end{itemize}

\subsection{Corner flow}

\begin{figure}[tbp]
        \centerline{\includegraphics{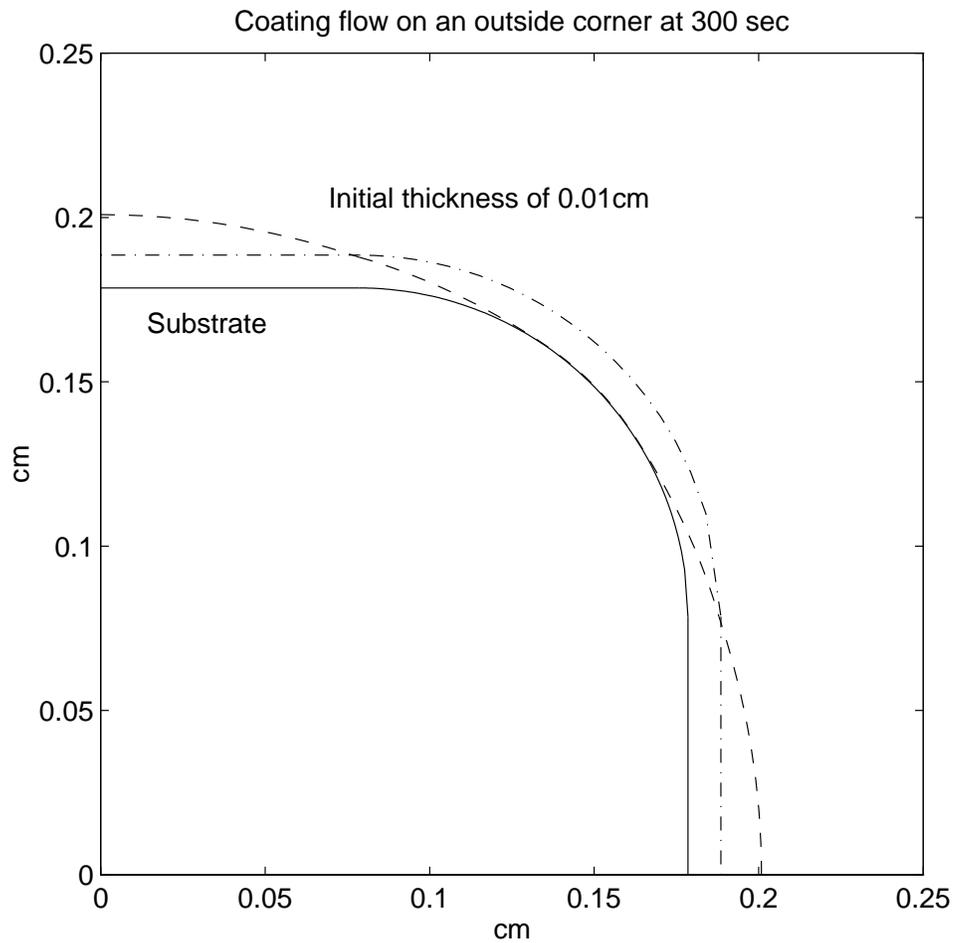}}
        \caption{The initial film of fluid (dot-dashed) of constant
        thickness around a corner (solid) evolves over a long time to thin
        around the corner (dashed).  (Unlike Schwartz \&
        Weidner~\protect\cite[Fig.7]{Schwartz95a}, the thickness of the
        film has \emph{not} been exaggerated.)}
        \protect\label{fcorner2}
\end{figure}

Consider here the two-dimensional fluid dynamics examined by Schwartz
\& Weidner~\cite{Schwartz95a} consisting of the flow which thins a
film around an outside corner as shown in Figure~\ref{fcorner2}.
The fluid is taken to have surface tension $\sigma=30\mbox{dynes/cm}$,
viscosity $\mu=1\mbox{poise}$, and density $\rho=1\mbox{gm/cm}^3$.
Initially the film is a uniform 0.01cm thick, corresponding to a
Reynolds number of $\cR=0.003$, around the corner which
has a radius of 0.1cm.  As in the previous subsection, the most
convenient way to parameterise the one-dimensional substrate is in
terms of the arclength $x$.  In this case our dynamical
model~(\ref{initstate}) reduces to
\begin{equation}
\D{\zeta }{t} =(1-\ka\eta)\D\eta t=
-  \rat{1}{3} \D\ x \left[\eta ^{2}\zeta \D\tka x \right]
+ \Ord{\ep^{5}}\,,  \label{lub1d}
\end{equation}
where $\tka = \ka + \ka^{2} \eta + \DD\eta x $, and $\ka=k_1(x)$ is
the curvature of the substrate.  In contrast, the model of Schwartz
\& Weidner (the SaW model) is
\begin{equation}
\D{\eta}{t} = - \third \D\ {x} \left[\eta^{3}
\D\ x\left(\ka + \DD\eta x\right) \right]\,.
\label{sawmod}
\end{equation}
The differences between the models are that ours includes more terms
in the curvature.  In particular, ours conserves fluid whereas the SaW
model does not.  Indeed, in the course of the numerical simulations of
the film flow shown in Figure~\ref{fcorner2}, the SaW model lost about
2\% of the fluid whereas ours lost none to computational error.
For thicker films the difference is more marked.

A numerical scheme to simulate a fluid film via these equations is
straightforward.  We approximated both~(\ref{lub1d}) and
(\ref{sawmod}) by finite differences on a spatial grid with $N=97$
points and, because the dynamics are stiff, we employed an implicit
integration scheme with time step $\Delta t=0.00115$s.  The numerical
scheme uses second-order centred differences in space and time but, in
the interests of speed, the nonlinear coefficients are only computed
from the earlier time.  By varying the size of the space-time grid we
determined that these parameters give a numerically accurate
simulation.

\begin{figure}[tbp]
       \centerline{\includegraphics{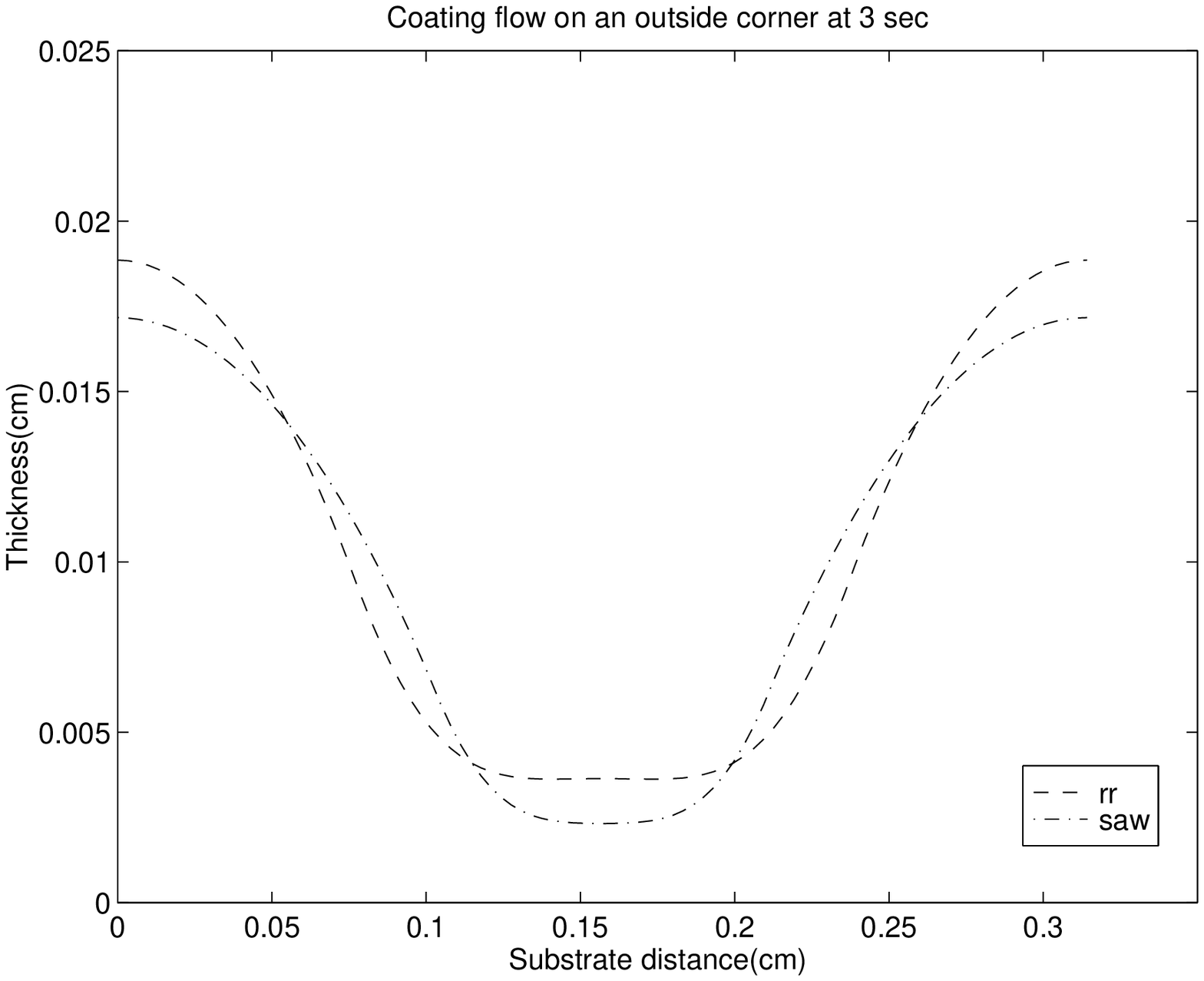}}
        \caption{film thickness for the flow around the corner shown in
        Figure~\protect\ref{fcorner2} at time $t=3$s: dashed is our
        model~(\protect\ref{lub1d}); dot-dashed is the SaW
        model~(\ref{sawmod}).}
        \protect\label{fcorn3}
\end{figure}

\begin{figure}[tbp]
       \centerline{\includegraphics{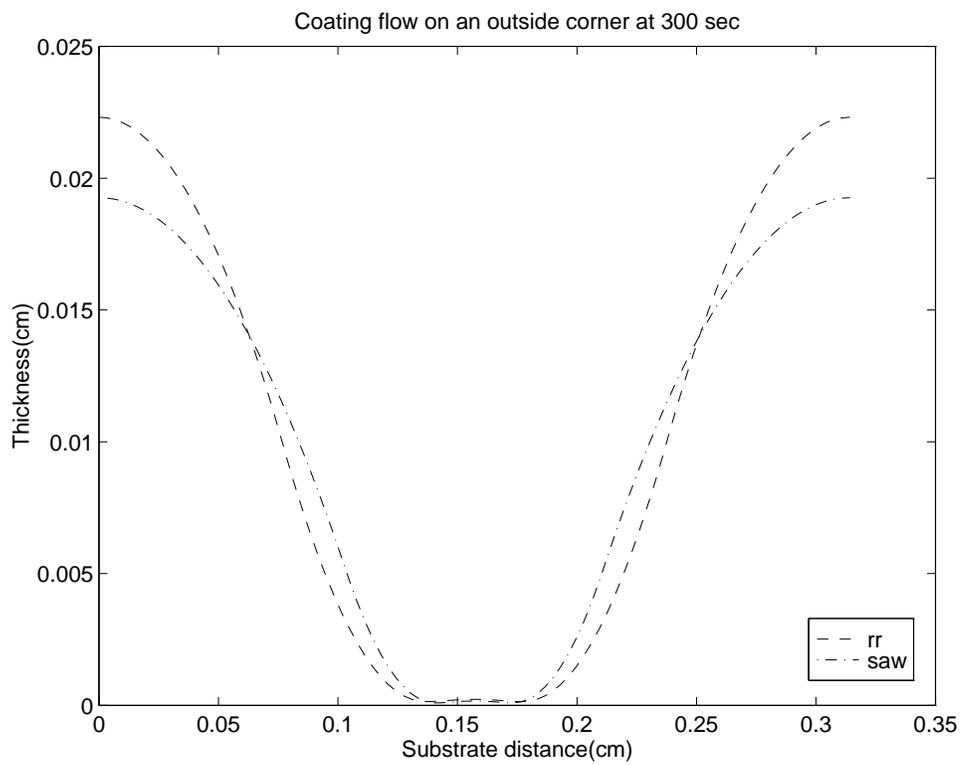}}
        \caption{film thickness for the flow around the corner shown in
        Figure~\protect\ref{fcorner2} at time $t=300$s: dashed is our
        model~(\protect\ref{lub1d}); dot-dashed is the SaW
        model~(\ref{sawmod}).}
        \protect\label{fcorn300}
\end{figure}

Shown in Figures~\ref{fcorn3} and~\ref{fcorn300} are comparisons
between the predictions of the SaW model and ours during the
simulation of the thinning of the film around the corner.  Observe
that the SaW model and ours are quantitatively different: the SaW
model predicts a more rapid thinning of the film around the corner,
and a slower thickening of the film away from the corner.  For
example, the thickness at $x=0$ and $t=3$s for our model is only
reached by the SaW model at time $t\approx 300$s.  For quantitative
accuracy the thin film flows need the extra curvature terms employed
in our model~(\ref{lub1d}).

Strictly, this corner flow is unsuitable for the lubrication
approximation because the sharp change in curvature from the flat wall
to the round corner generates arbitrarily large gradients of the
curvature $\kappa$.  These gradients are compounded at higher order.
For example, a computation of the Reynolds number terms in the first
line of~(\ref{eqrey}) for this problem reveals a relative magnitude,
when compared to $\partial\eta/\partial t$, of typically
$1000\cR\approx 3$.  Thus such high-order effects may be significant
near the sharp change in curvature.  In practise, perhaps we should
view the sharp change as an element of roughness in an otherwise
smooth transition between the wall and the round corner.
Alternatively, we could view it as a discontinuity between two smooth
(constant) curvature domains and thereby seek matching conditions via
similar arguments to those employed in~\cite{Roberts92c} for boundary
conditions.

\subsection{Flow around a torus}
\label{Storus}

Here we discuss the evolution of the flow over the surface of a torus
with tube of radius $R_2$ and centreline of radius $R_1$.  We use the
following parameterisation:
\begin{equation}
\begin{array}{l}
X_{1} = (R_{1}+R_{2}\cos\theta ) \cos\phi \,,\\
X_{2}= (R_{1}+R_{2}\cos\theta ) \sin\phi \,,\\
X_{3}= R_{2} \sin\theta\,.
\end{array}
\end{equation}
Take $R_{2} < R_{1}$ to avoid self-intersection, and denote the
coordinates $x_{1}= \phi$ and $x_{2}=\theta$ with $0 \leq \phi < 2\pi$
and $0 \leq \theta < 2\pi$.
Then this parameterisation generates an
orthogonal curvilinear coordinate system with corresponding
orthonormal basis vectors:
\begin{equation}
\begin{array}{l}
\ex=\vec e_{\phi}= - \sin\phi \vec i +\cos\phi \vec j \,,  \\
 \ez=\vec e_{\theta}= - \sin\theta ( \cos\phi \vec i +
 \sin\phi \vec j) + \cos\theta \vec k \,,  \\
 \ey=\ex \times \ez =\cos\theta(\cos\phi\vec i +\sin\phi\vec j)
 +\sin\theta \vec k\,,
\end{array}
\end{equation}
in terms of the unit vectors of the conventional Cartesian
coordinate system, and with the corresponding surface metrics
\begin{equation}
        \hsx=m_{\phi}=  R_{1}+ R_{2} \cos\theta \,,  \quad
        \hsz=m_{\theta}= R_2\, .
\end{equation}
The chosen coordinate system also generates lines of principal
curvature as the parameter curves.  We then obtain the following
principal curvatures and mean curvature:
\begin{equation}
k_{1}=  - {\cos\theta  \over  R_{1} + R_{2} \cos\theta  } \, ,
\quad
k_{2}= - {1\over R_{2}}  \, ,
\quad
\kappa = - {1\over R_{2}}  {R_{1} + 2 R_{2} \cos\theta \over
R_{1} + R_{2} \cos\theta}\,.
\end{equation}

The mean curvature of the substrate is
maximum at the inner rim of the torus ($\theta = \pi$), and minimum at
the outer rim ($\theta = 0$).  Hence we expect the fluid layer to thicken
around the inner rim, and to thin around the outer, solely due to
surface tension effects.

We numerically solve the model~(\ref{lubeps}).  For simplicity, we
seek axisymmetric solutions, that is solutions independent of the
angle $\phi$ around the rim of the torus, and so the film thickness
depends only upon $\theta$, the angle around the tube, and $t$.  We
try the following form for $\eta$
\begin{equation}
        \eta (t, \theta) = \sum_{n=0}^{N-1} a_{n}(t) \cos(n\theta)\,,
\end{equation}
which guarantees the periodicity of the solution and imposes symmetry
across the plane $z=0$.  The ordinary differential equations for the
coefficients $a_n$ are found by a Galerkin method where they are
determined by making the corresponding residual error orthogonal to
the $N$ basis functions, $\cos(n\theta)$, in the usual $L_{2}$ norm.
As a check, we confirm that the total volume enclosed between the
free surface of the fluid and the toroidal substrate remains constant
in time.  Two numerical simulations were done.
\begin{itemize}
\item First, we start with an initially uniform layer of thickness
$\eta_{0}=0.1$ on a torus with $R_{1}=2$ and $R_{2}=1$.  The
corresponding flow towards the inner rim of the torus is shown by the
evolution of the film thickness in Figure~\ref{ftunif}.
\begin{figure}[tbp]
        \centerline{(a) \includegraphics{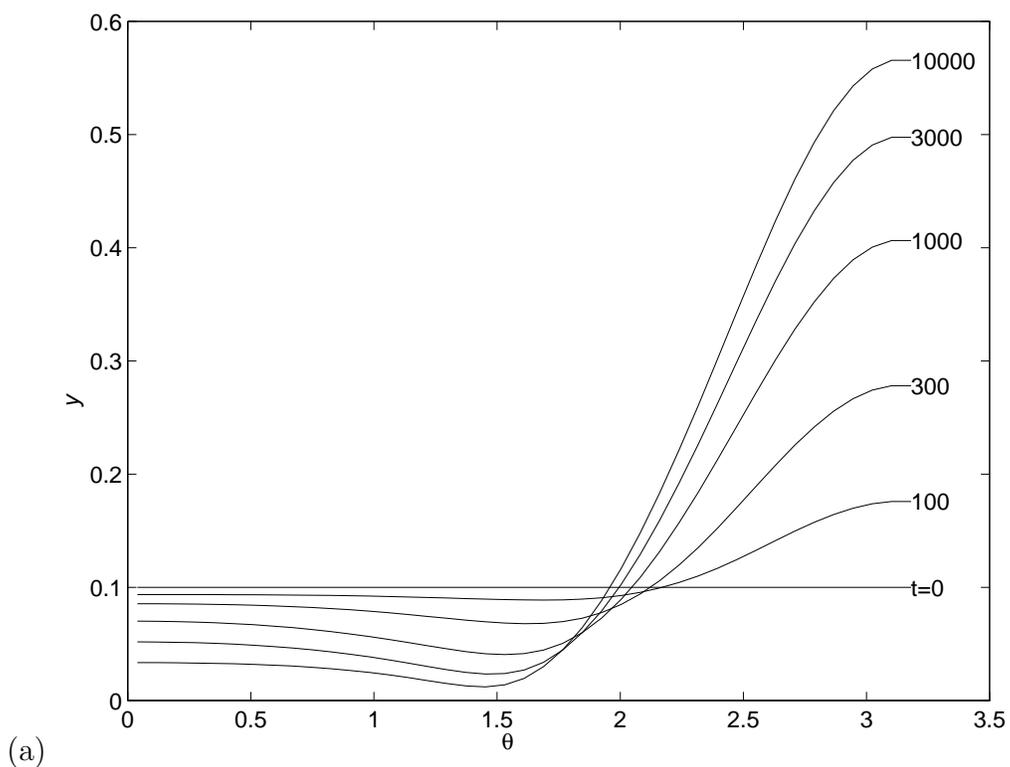}}
        \centerline{(b) \epsfxsize=75mm\includegraphics{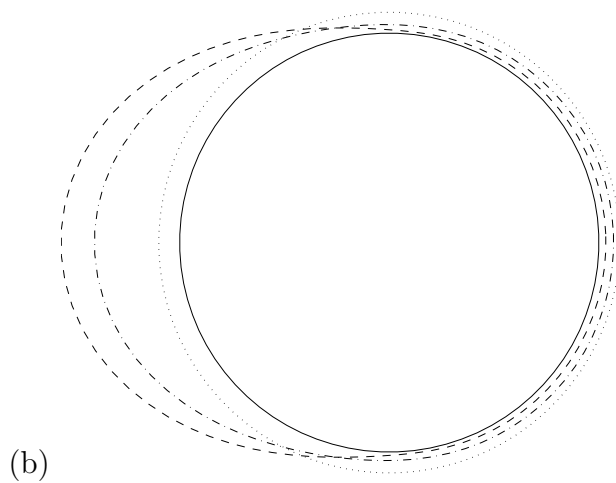}}
        \caption{evolution of flow on the surface of a torus with $N=15$
        terms in Galerkin approximation.  $R_{1}=2$, $R_{2}=1$, $\eta_{0}=
        0.1$: (a) $\eta (t, \theta)$ shown at $t=0$, 100, 300, 1000, 3000,
        10000; (b) initial, intermediate ($t=1000$) and late ($t=10000$)
        stages of the film on a cross-section of the torus.}
        \protect\label{ftunif}
\end{figure}

\item Second, we simulated the flow evolving from a strip of fluid
placed around the outer rim of the torus.  Again, as shown in
Figure~\ref{ftstep}, the fluid flows around to the inner rim.
\begin{figure}[tbp]
        \centerline{\includegraphics{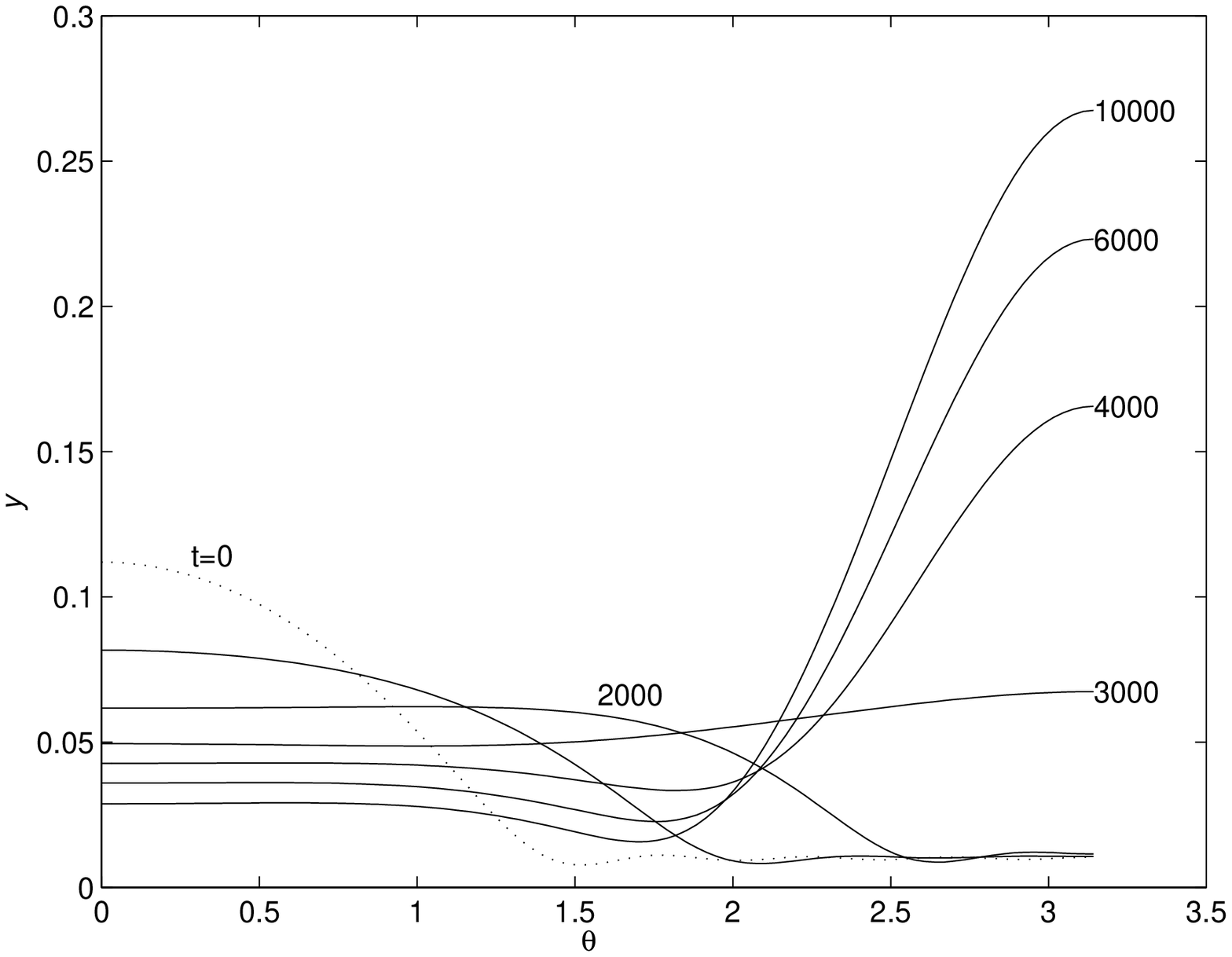}}
        \caption{Same as Figure~\protect\ref{ftunif} with a step-like
        initial layer (dotted).  $N=15$ and $t=0$, 1000, 2000, 3000, 4000,
        6000, 8000, 10000.}
        \protect\label{ftstep}
\end{figure}

\end{itemize}

On the torus, the importance of inertia effects on this lubrication
flow are estimated by the ratio between typical values of the
right-hand side of~(\ref{lubeps}) and that of~(\ref{eqrey}):
dominantly
\begin{displaymath}
        \mbox{inertia}:\mbox{lubrication}
        \approx\frac{\cR \eta^4}{6R_1R_2^2}\,.
\end{displaymath}
As may be expected, a thicker film or a more sharply curved torus are
more likely to be affected by such higher-order influences. Note that
for flows inside a toroidal tube, the thinning of the liquid layer
occurs around the inner rim ($\theta = \pi$). Three-dimensional simulations
of coating flows exhibiting transversal instabilities (along the
$\phi$-direction)
similar to those found in
\cite{Weidner97} will be published elsewhere.

\paragraph{Acknowledgements:} RVR would like to acknowledge the
many helpful discussions with L.W.~Schwartz.
His work was completed while spending a sabbatical leave at
USQ whose warm hospitality is also gratefully acknowledged.

\end{document}